\documentclass[12pt,preprint]{aastex}

\lefthead{LEE ET AL.}
\righthead{LOOK-BACK TIME EVOLUTION OF FAR-UV FLUX FROM ELLIPTICAL GALAXIES}

\begin{document}

\title{THE LOOK-BACK TIME EVOLUTION OF FAR-UV FLUX FROM ELLIPTICAL GALAXIES:\\ THE FORNAX CLUSTER AND ABELL~2670}

\author{
Young-Wook Lee\altaffilmark{1},
Chang H. Ree\altaffilmark{1},
R. Michael Rich\altaffilmark{2},
Jean-Michel Deharveng\altaffilmark{3},
Young-Jong Sohn\altaffilmark{1},
Soo-Chang Rey\altaffilmark{1, 4},
Sukyoung K. Yi\altaffilmark{5},
Suk-Jin Yoon\altaffilmark{5},
Luciana Bianchi\altaffilmark{6}, 
Jae-Woo Lee\altaffilmark{7},
Mark Seibert\altaffilmark{4},
Tom A. Barlow\altaffilmark{4},
Yong-Ik Byun\altaffilmark{1},
Jose Donas\altaffilmark{3},
Karl Forster\altaffilmark{4},
Peter G. Friedman\altaffilmark{4},
Timothy M. Heckman\altaffilmark{8},
Myungkook J. Jee\altaffilmark{8},
Patrick N. Jelinsky\altaffilmark{9},
Sug-Whan Kim\altaffilmark{1},
Barry F. Madore\altaffilmark{10,11},
Roger F. Malina\altaffilmark{3},
D. Christopher Martin\altaffilmark{4},
Bruno Milliard\altaffilmark{3},
Patrick Morrissey\altaffilmark{4}, 
Susan G. Neff\altaffilmark{12},
Jaehyon Rhee\altaffilmark{1, 4},
David Schiminovich\altaffilmark{4},
Oswald H. W. Siegmund\altaffilmark{9}, 
Todd Small\altaffilmark{4},
Alex S. Szalay\altaffilmark{8}, 
Barry Y. Welsh\altaffilmark{9}, and
Ted K. Wyder\altaffilmark{4}
}

\altaffiltext{1}{Center for Space Astrophysics and Department of Astronomy, Yonsei University, Seoul 120-749, Korea (e-mail : ywlee@csa.yonsei.ac.kr, chr@csa.yonsei.ac.kr)}
\altaffiltext{2}{Department of Physics and Astronomy, University of California at Los Angeles, Los Angeles, CA 90095}
\altaffiltext{3}{Laboratoire d'Astrophysique de Marseille, BP 8, Traverse du Siphon, 13376 Marseille Cedex 12, France}
\altaffiltext{4}{California Institute of Technology, MC 405-47, 1200 East California Boulevard, Pasadena, CA 91125}
\altaffiltext{5}{Department of Physics, University of Oxford, Oxford OX1 3RH, UK}
\altaffiltext{6}{Center for Astrophysical Sciences, Johns Hopkins University, 3400 North Charles St., Baltimore, MD 21218}
\altaffiltext{7}{Department of Astronomy and Space Sciences, Sejong University, Seoul 143-747, Korea}
\altaffiltext{8}{Department of Physics and Astronomy, The Johns Hopkins University, Homewood Campus, Baltimore, MD 21218}
\altaffiltext{9}{Space Sciences Laboratory, University of California at Berkeley, 601 Campbell Hall, Berkeley, CA 94720}
\altaffiltext{10}{Observatories of the Carnegie Institution of Washington, 813 Santa Barbara St., Pasadena, CA 91101}
\altaffiltext{11}{NASA/IPAC Extragalactic Database, California Institute of Technology, Mail Code 100-22, 770 S. Wilson Ave., Pasadena, CA 91125}
\altaffiltext{12}{Laboratory for Astronomy and Solar Physics, NASA Goddard Space Flight Center, Greenbelt, MD 20771}

\begin{abstract}
In order to investigate the origin of the far-UV (FUV) flux from the early-type galaxies, Galaxy Evolution Explorer (GALEX) is collecting the UV data for the elliptical-rich clusters at moderate redshifts ($z < 0.2$) where the dominant FUV source is predicted to be hot horizontal-branch (HB) stars and their post-HB progeny. Here we present our first result for the early-type galaxies in Abell~2670 at $z = 0.076$. Compared to NGC~1399, a nearby giant elliptical galaxy in the Fornax cluster, it appears that the rest-frame $FUV - V$ color of the giant ellipticals gets redder by $\sim$ 0.7 mag at the distance of Abell~2670 ($z = 0.076$; look-back time $\approx$ 1.0 Gyr). Although a detailed comparison with the models is postponed until more cluster data are accumulated, it is interesting to note that this value is consistent with the variation predicted by the population synthesis models where the mean temperature of HB stars declines rapidly with increasing look-back time.
\end{abstract}


\section{INTRODUCTION}
It is now well established that the far-UV (FUV) flux (``UV upturn'') of nearby early-type galaxies originates from a minority population of old hot helium-burning horizontal-branch (HB) stars (e.g., O'Connell 1999; Brown et al. 2000b). Stellar evolution models of HB predict that the mean temperature of HB distribution declines rapidly with decreasing age (e.g., Lee et al. 1994), and therefore the FUV flux from ellipticals should fade rapidly with look-back time (e.g., Greggio \& Renzini 1990; Tantalo et al. 1996). Population synthesis models of Yi et al. (1999) also indicate that careful observations for the UV look-back time evolution could also discriminate two alternative HB solutions on the origin of UV flux, i.e., ``metal-poor'' and ``metal-rich'' HB models, which predict different ages for giant ellipticals. 

The $Hubble~Space~Telescope$ ($HST$) has obtained FUV images for several elliptical-rich clusters in the redshift range of $0.3 < z < 0.6$ (Brown et al. 2000a, 2003), but stellar evolution models (Lee et al. 1999; Yi et al. 1999) predict that hot HB stars must be absent at these relatively high redshifts, as no stars of low enough mass have yet evolved. Therefore, UV observations covering a lower redshift range ($z < 0.2$) are still required, in order to test the look-back time evolution effect of HB stars in giant ellipticals. Here we present the first result from the Galaxy Evolution Explorer (GALEX) FUV observations for one Abell cluster in this redshift range, and also for the Fornax cluster as a local calibrator.

\section{OBSERVATIONS AND DATA REDUCTIONS}

%place fig1 here

GALEX imagery of Abell~2670 (hereafter A2670) at $z = 0.076$ is based on 8 orbits (8,309~seconds) of exposure. These are the first observations for this study, and were obtained on 15-16 September 2003. The GALEX images, in the near-UV ($1750 - 2750$~\AA) and far-UV ($1350 - 1750$~\AA) bandpasses, were centered at R.A. = 23$^h$54$^m$10$^s$ and DEC. = -10\arcdeg24\arcmin18\arcsec~(J2000). The photon count maps from the single visits were coadded to make the image for photometry, and it contains 3840~$\times~$3840 pixels with 1.5~arcsec/pixel scale. The  calibrations and preprocessings were performed by the GALEX pipeline. In order to compare the remote targets with the local samples, we have also imaged the Fornax cluster of galaxies centered at R.A. = 3$^h$38$^m$30$^s$ and DEC. = -35\arcdeg30\arcmin18\arcsec~(J2000). The exposure time for the Fornax field was 1,704~seconds (1 orbit), and the UV images were also processed through the GALEX pipeline in the same manner. The GALEX data pipeline utilizes the SExtractor image analysis package (Bertin \& Arnouts 1996) for detection and photometry of the sources in the imaging data with some modifications in the determinations of sky background and detection threshold (Martin et al. 2004; Morrissey et al. 2004).

For the identification and optical photometry of the galaxies in A2670, we used $B$ and $r$ images from Fasano et al. (2002). They have classified 33 early-type galaxies within a 6.5~$\times$~6.5~arcmin$^2$ field, and among the member galaxies confirmed by redshift values (see their Table 7g), we have identified 10 galaxies (5 E's and 5 S0's) in the FUV image with 3$\sigma$ detection threshold (Figure~1). We obtained the optical images of A2670 from the $Nordic~Optical~Telescope$ (NOT, La Palma) archive, and ran the SExtractor photometry package with default parameters. The $V$ magnitudes were estimated from $B$ and $r$ images by adopting the relation, $(V - r) = 0.36~(B - r) - 0.21$, from Kent (1985). We confirmed this relation agrees well with the measured early-type galaxy colors at $z = 0.0 \sim 0.1$ in Frei \& Gunn (1994) to within $\sim$ 0.05 mag.

In order to compare the local sample with the remote targets without any systematic effect that may arise from the different aperture sizes used in the photometry, we have measured total magnitudes for the early-type galaxies in both A2670 and the Fornax cluster. In practice, we have selected MAG$\_$AUTO measurements from the elliptical aperture photometry in the SExtractor catalogs for both FUV and $V$ bandpasses as the total magnitudes for the galaxies in A2670. Galactic foreground extinctions were then corrected with $R_V = 3.1$, $R_{FUV} = 8.16$ (Cardelli et al. 1989), and $E(B - V)$ = 0.043 (Schlegel et al. 1998). FUV magnitudes for the early-type galaxies in the Fornax cluster were measured from the surface photometry of GALEX FUV image. The FUV surface brightness profiles extend to $r > 80\arcsec$ in the case of NGC~1399 and NGC~1404 (see Figure~2), and we calculate the FUV total magnitudes by integration of the radial profiles. We have also corrected for the Galactic foreground extinctions for the galaxies in the Fornax cluster ($A_{FUV}$ = 0.09 $\sim$ 0.14), and their total magnitudes in $V$ band were adopted from the Third Reference Catalog of Bright Galaxies (RC3; de Vaucouleurs et al. 1991). Our FUV photometry for the elliptical galaxies in the Fornax cluster, when measured within the $International~Ultraviolet~Explorer~Satellite$ ($IUE$) aperture (20\arcsec $\times$ 10\arcsec), agrees well with the $IUE$ measurements (Burstein et al. 1988) to within 0.01 $\sim$ 0.06 mag.

%place fig2 here

\section{RESULTS AND DISCUSSION}

For the cross-matched early-type galaxies in the Fornax cluster and A2670, the apparent extinction-corrected total magnitudes, in AB magnitude system (Oke 1974), are plotted in Figure~3 for both FUV (filled symbols) and $V$ (open symbols) bandpasses. The redshift evolution predicted by the population synthesis models are also compared under two alternative assumptions on the metallicity of FUV source. The ``metal-poor HB model'' suggests that the dominant FUV sources are very old ($t > 12$ Gyr) hot metal-poor HB stars and their post-HB progeny, while the ``metal-rich HB model'' suggests that the dominant sources are old ($t \sim 10 - 12$ Gyr) super metal-rich ([Fe/H] $>$ 0.0) hot HB stars that experienced enhanced helium enrichment and mass-loss (see Yi et al. 1999). The models were constructed to match the $FUV - V$ color of the local giant elliptical galaxy NGC~1399, the brightest galaxy in the Fornax cluster, and then passively evolved with look-back time. The models are then redshifted and the look-back times are converted to the equivalent redshift values, adopting the currently favored cosmological parameters ($\Omega_M$, $\Omega_\Lambda$, $H_0$) = (0.3, 0.7, 70). Therefore, the foreground extinction-corrected observed data can be directly compared with the models.  Since the NGC~1399 is a typical FUV strong giant elliptical galaxy in the local Universe, the model lines would represent the redshift (look-back time) evolution of giant ellipticals, and hence should be compared with the brightest ellipticals in each cluster.

The models presented here are similar to those in Yi et al. (1999), except that Lejeune et al. (1997) theoretical spectral library is employed, and a fixed mass of 0.565 $M_\odot$ is assumed as the typical value of post-asymptotic giant branch (PAGB) mass for the progenies of the main-sequence stars with $M_\odot$ = 0.8 $\sim$ 1.0. The reader is referred to Yi et al. (1999) for the details of model constructions.

In Figure~3, we have also plotted the $HST$ data for the early-type galaxies in two rich clusters, CL 1358+62 at $z = 0.33$ and CL 0016+16 at $z = 0.55$. These clusters were observed with the Space Telescope Imaging Spectrograph (STIS) in FUV by Brown et al. (2000a, 2003) and with the Wide Field Planetary Camera 2 (WFPC2) in optical bandpasses by van Dokkum et al. (1998; CL 1358+62) and Smail et al. (1997; CL0016+16). We have taken their photometric measurements in FUV (F25QTZ) and $V$ (F555W or F606W) bandpasses from the literature, and converted those $HST$ magnitudes to GALEX AB magnitude system using the photometric zeropoints and pivot wavelengths defined in the $HST$ image headers ($m_{AB} = m_{ST} - 5 \log\lambda_p + 18.6921$). Model spectral energy distributions (SEDs) redshifted to $z$ = 0.33 and 0.55 were used to estimate the systematic difference between the $HST$ and GALEX filter systems. Since the aperture radii in their analyses (0$\farcs$4 in FUV for both clusters, and 1$\farcs$5 in $V$ for CL 1358+62) are not large enough to represent total magnitudes, we applied appropriate aperture corrections, -0.32 (FUV) and -0.22 ($V$) for CL 1358+62. For the galaxies in CL 0016+16, we applied a correction only in FUV (-0.17 mag), as we adopted the total $V$ magnitude from the catalog in Smail et al. (1997). The aperture correction values were obtained by utilizing the FUV (this paper) and $V$ (Marcum et al. 2001) radial profiles of NGC~1399, assuming that these clusters are at 75 (CL 1358+62) and 117 (CL 0016+16) times the distance of the Fornax cluster with the assumed cosmological parameters above. Finally, foreground extinction corrections, $E(B - V)$ = 0.023 and 0.057 (Schlegel et al. 1998), are applied for CL 1358+62 and CL 0016+16, respectively, to obtain the apparent extinction-corrected total magnitudes in FUV and $V$ for the galaxies in these clusters. The $HST$ Faint Object Camera data for Abell~370 at $z = 0.375$ (Brown et al. 1998) are not used here, as they may have significant systematic errors according to Brown et al. (2003).

%place fig3 here

Some early-type galaxies show contamination from the residual star formation in their UV spectra (Burstein et al. 1988; Yi et al. 2004), and therefore it is important to check whether our sample galaxies are not affected by a minority population of young stars. First, we have checked GALEX near-UV (NUV) photometry, as the NUV light is very sensitive to the presence of young main-sequence stars (Yi et al. 2004). In A2670, only one S0 galaxy, a15, shows exceptionally strong NUV flux, while other member galaxies follow the general correlation on the $FUV - V$ vs. $NUV - V$ diagram. We have then searched the Sloan Digital Sky Survey (SDSS) database to inspect the optical spectra for member galaxies in A2670. Among the four galaxies (a10, a16, a19, and a21) matched in SDSS spectroscopy database, one (a16) shows H$\alpha$ emission, while others show no obvious emission features. As noted by Brown et al. (2003), CL 1358+62 has a FUV bright galaxy (375), which is contaminated by extended Ly$\alpha$ emission. We also found that a galaxy in CL 0016+16, DG269, shows unusually strong dependence of $FUV - V$ color on the aperture size used in the photometry, and is not likely a normal quiescent elliptical galaxy. All of these abnormal galaxies with recent star-forming signatures are excluded in our analysis. The reasonable agreement of the models with the brightest quiescent galaxies in Figure~3 would confirm that we are actually comparing the passively evolving systems.

From the total magnitudes measured in both bandpasses, we have derived $FUV - V$ colors for the normal quiescent galaxies in Figure~3. The apparent redshift evolution of the observed $FUV - V$ colors is then presented in Figure~4, along with the model predictions. As in Figure~3, the models are for the giant ellipticals in clusters, and therefore only the FUV strong ellipticals in each cluster (e.g., a19 in A2670) should be compared with the models. It is clear from Figure~4 that we have detected the fading of the UV upturn expected at moderate redshifts. Compared to NGC~1399, a nearby giant elliptical galaxy in the Fornax cluster, an apparent extinction-corrected $FUV - V$ color of a giant elliptical galaxy (a19) in A2670 gets redder by $\sim$ 0.55 mag at the distance of A2670 ($z = 0.076$; look-back time $\approx$ 1.0 Gyr). For the rest-frame $FUV - V$ color, this corresponds to 0.70 mag, which is consistent with the variation predicted by the models. In these models, the  dominant FUV source is hot HB stars for $z < 0.25$, and therefore the FUV flux fades rapidly with look-back time (redshift) as the HB temperature distribution becomes cooler with decreasing age (increasing look-back time). 

%place fig4 here

For $z > 0.3$, the $HST$ data also appear to be in reasonable agreement with our model predictions, to within the errors. Note, however, that the models become more uncertain at these relatively higher redshifts, as the dominant FUV source changes from hot HB stars to PAGB stars for $z > 0.25$. In particular, Lee et al. (1999, see their Figure~1) predict that the HB contribution to the total FUV flux becomes almost negligible for $z > 0.4$. At relatively higher redshifts, the total FUV flux would therefore increase as the mass of PAGB stars assumed decreases because their lifetimes correlate inversely with their mass. Consequently, although the available $HST$ data may provide no direct test of HB evolution effect, they nevertheless provide a useful test on the adopted mass (0.565 $M_\odot$) of PAGB stars in our models. 

With only two clusters at relatively low redshifts, it is premature to conclude that GALEX has detected the look-back time evolution of the mean temparature of HB stars in early-type galaxies. It is interesting to note, however, that the observed fading of the UV upturn is consistent with the variation predicted by the population synthesis models where the mean temperature of HB stars declines rapidly with increasing look-back time. Further observations for more cluster targets are clearly needed to confirm the universality of this trend, and also to attempt a more detailed comparison between the two alternative models on the metallicity of hot HB stars, and with other possible models for the origin of the UV upturn. Planned observations with GALEX to similar depth for about a dozen of elliptical-rich clusters at moderate redshifts ($0.0 < z < 0.20$) are highly anticipated in this regard.

\acknowledgments{
GALEX (Galaxy Evolution Explorer) is a NASA Small Explorer, launched in April 2003. We gratefully acknowledge NASA's support for construction, operation, and science analysis for the GALEX mission, developed in cooperation with the Centre National d'Etudes Spatiales of France and the Korean Ministry of Science and Technology. Yonsei University participation is funded by the Korean Ministry of Science \& Technology, for which we are grateful. JWL also acknowledges support from KOSEF to the ARCSEC.}

\begin{figure}
\epsscale{0.5}
\plotone{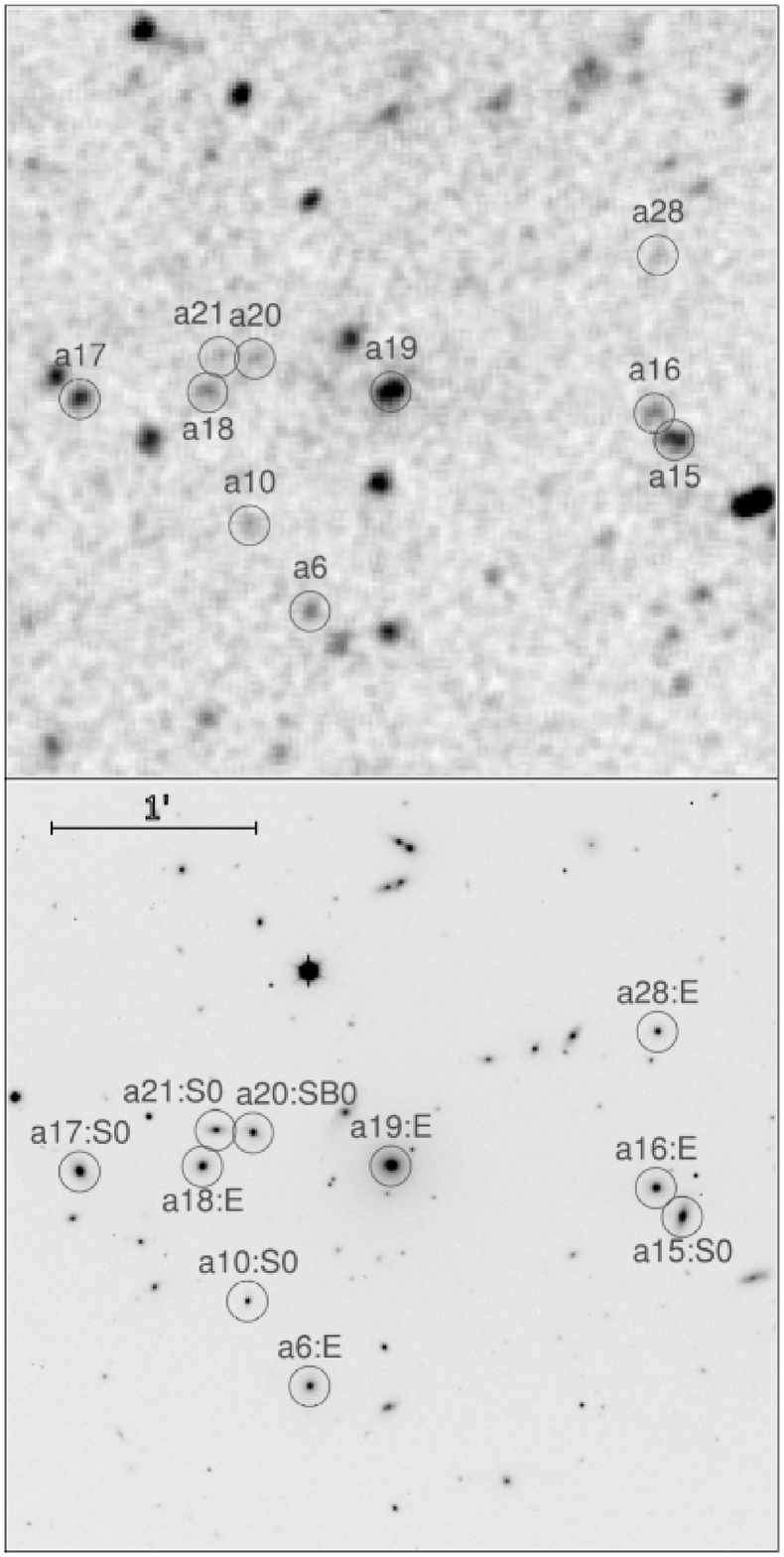}
\caption{5~$\times~$5~arcmin$^2$ portion of the GALEX FUV ($top$) and optical ($bottom$) images centered on the Abell~2670. Cross-matched early-type galaxies are marked by the 7$\farcs$5 radius circles.\label{fig1}}
\end{figure}

\begin{figure}
\epsscale{0.5}
\plotone{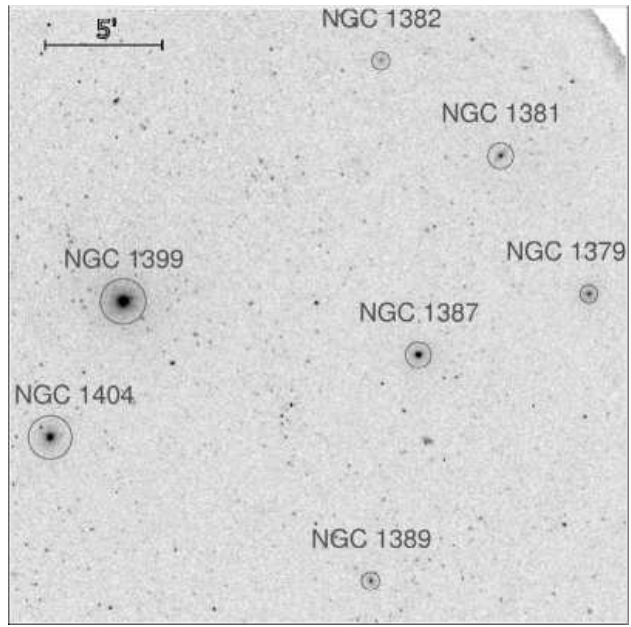}
\caption{40~$\times$40~arcmin$^2$ portion of the GALEX FUV image for the Fornax cluster. Cross-matched early-type galaxies are identified with the circles.\label{fig2}}
\end{figure}

\begin{figure}
\epsscale{1.0}
\plotone{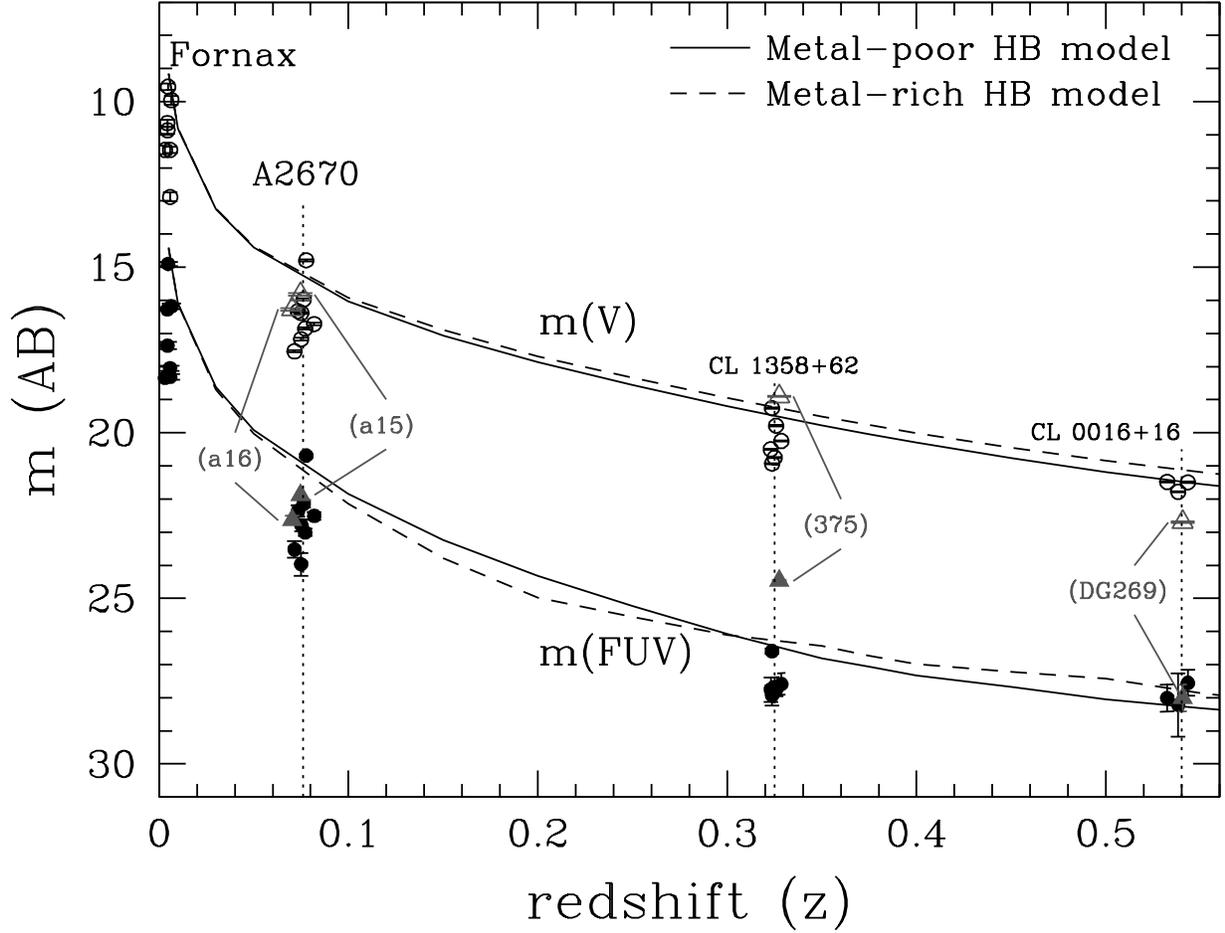}
\caption{The redshift evolution of the apparent extinction-corrected total AB magnitudes for the early-type galaxies in the Fornax cluster and Abell~2670. The FUV magnitudes were measured from the GALEX images. The $HST$ data for two rich clusters at $z > 0.3$, CL 1358+62 and CL 0016+16, are also plotted. The solid and dashed lines are the predictions from the passively evolving population synthesis models for the giant elliptical galaxies, where the effects of stellar evolution and redshift are all included (see the text). The triangles represent the galaxies with star-forming signatures.\label{fig3}}
\end{figure}

\begin{figure}
\epsscale{1.0}
\plotone{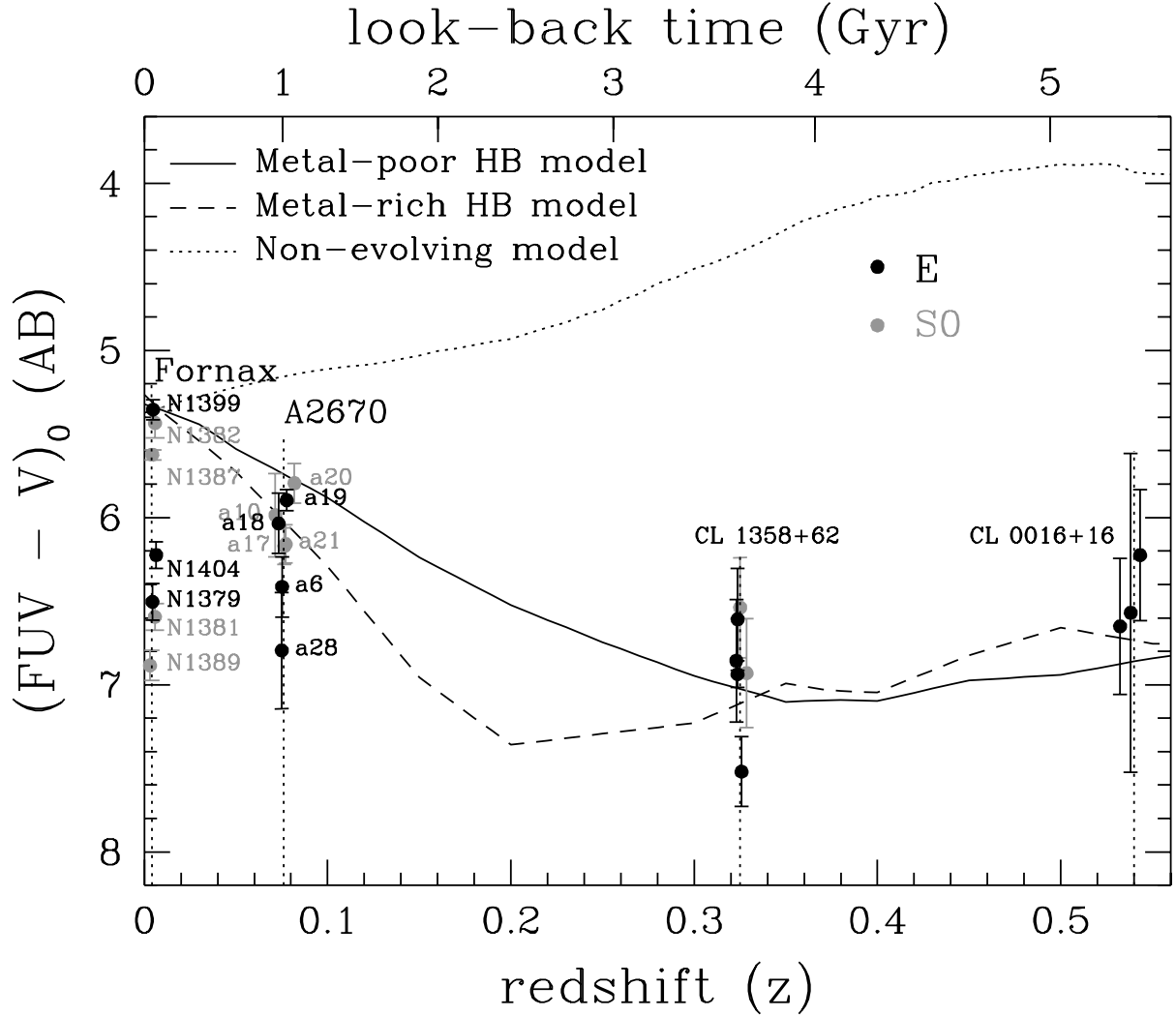}
\caption{Similar to Figure~3, but for the redshift evolution of $FUV - V$ color for the quiescent early-type galaxies in sample clusters. Note that the color of a giant elliptical galaxy (a19) in A2670 fades by ~0.55 mag. compared to the NGC~1399, a nearby giant elliptical galaxy in the Fornax cluster. The solid and dashed lines are two alternative models on the FUV evolution of normal giant ellipticals, while the dotted line is for the case with only the effect of redshift and without the effect of stellar evolution.\label{fig4}}
\end{figure}

\end{document}